\documentclass[12pt]{iopart}
\usepackage{iopams}
\usepackage{braket}
\usepackage{graphicx}
\usepackage{cite}
\def\nuc#1#2{$^{#1}$#2}

\def\scala{0.45}
\def\scalb{0.5}

\def\beqn{\begin{equation}}
\def\eeqn{\end{equation}}
\def\bpmax{\left(\begin{array}{cc}}
\def\epmax{\end{array}\right)}
\def\Bbb{B^{}_{\B\B}}
\def\B{\beta}
\def\Bbg{B^{}_{\B\G}}
\def\Bgg{B^{}_{\G\G}}
\def\clB{{\cal B}}
\def\dbe{\partial_{\B}}
\def\dga{\partial_{\G}}
\def\G{\gamma}
\def\Relp{{\rm Re}}
\def\rnw{\sqrt{r\over w}}
\def\Sg{\sin\!3\G}
\def\we#1{\boldsymbol #1}

\begin{document}
\title{Microscopic description of collective properties of even-even Xe isotopes}
\author{L Pr\'ochniak}
\address{Heavy Ion Laboratory, University of Warsaw\\ Pasteura 5a, 02-093
Warsaw, Poland}
\begin{abstract}
Collective properties of the even-even \nuc{118-144}{Xe} isotopes have been
studied within a model employing the general Bohr Hamiltonian derived from
the mean-field theory based on the UNEDF0 energy functional. The calculated
low energy spectra and E2 transition probabilities are in good agreement
with experimental data.
\end{abstract}



\submitto{\PS}

\section{Introduction}

Currently developed mean-field theories based on energy density functionals
ambitiously aim to properly explain a large range of nuclear properties and
in many fields they have been quite successful. In this paper I present the
results of applying of the UNEDF0 functional \cite{2010KO29} to describe
low-energy collective excitations in the chain of even-even
\nuc{118-144}{Xe} isotopes. The treatment of collective properties is based
on the Adiabatic Time Dependent HFB (ATDHFB) theory, which leads to a
construction of a collective Hamiltonian from a microscopic, mean-field,
input. More details of the applied methods can be found in
\cite{2004PR01,2009PR08}. An alternative approach to collective phenomena
within a microscopic theory with phenomenological interactions employs the
Generator Coordinate Method (GCM), see e.g.
\cite{2010RO20,2013YA05,2008BE29}. One should also mention an attempt to
describe collective phenomena, in the RPA context however, starting
form realistic interactions and using the Unitary Correlation Operator
method (UCOM) \cite{2014PA10,2007PA08}.

For many years Xe nuclei have been subject of extensive experimental and
theoretical studies. Many of them were focused on a phenomenon of the double
$\B$ decay (confirmed experimentally in the case of \nuc{136}{Xe}) but there
is also a rich literature concerning the collective properties related to
changes of nuclear deformation. The long chain of Xe isotopes offers a good
opportunity to study the evolution of these properties as the number of
neutrons increases as well as the role of nonaxiality. Let me mention only a
few works which studied one or more of the even-even Xe isotopes: papers
employing geometrical concepts and the Bohr Hamiltonian
\cite{1977RO30,1999PR03,2004PR01,2012HI02,2013BO24}, papers based on the
interacting boson model \cite{1988VO09,1991KR10,2008SO12,2011CO07}, papers
using truncated shell model space \cite{1996PA03,2008ME04,2011HI05}.

\Sref{sec:teor} presents some basic facts on the general
Bohr Hamiltonian, on methods which allow for its derivation from the
mean-field theory and on the UNEDF0 density energy functional. In
\sref{sec:res} I show the results of calculations concerning several
 low-spin energy levels, some E2 transitions in the
\nuc{118-144}{Xe} nuclei as well as the comparison with experimental data.

\section{Theory\label{sec:teor}}
\subsection{Quadrupole variables, the Bohr Hamiltonian}

A consistent description of nuclear vibrational and rotational excitations as
well as of possible couplings between them requires the use of quadrupole collective variables i.e. of the
second rank with respect to the SO(3) rotation group.
Such variables can be chosen in various ways, e.g. as
parameters describing the shape of a nucleus \cite{x52bo01,x69bor} or the shape of
a phenomenological one-particle potential \cite{1976KA30,1977RO30,1999PR03}.
Within a self-consistent mean field theory such quadrupole variables
$\alpha_\mu$ (in the
laboratory frame) are chosen so as to be proportional to the components of the quadrupole
mass tensor
\begin{equation} \alpha_{2\mu}\sim
\langle \Phi |\sum_{i=1}^A r_i^2 Y_{2\mu}(\theta_i,\phi_i) |\Phi\rangle 
\end{equation} 
where 
 $\Phi$ is a microscopic nuclear wave function which 
can be obtained by using effective interactions of the Skyrme
\cite{2004PR01} or the Gogny type \cite{1999LI38,2010DE02} or in the  
relativistic framework (RMF) \cite{2004PR03,2009NI04}.  
The quadrupole variables can be equivalently expressed in the intrinsic
frame
(also called principal axes frame) by two deformation variables $\B$, $\G$
and three Euler angles ($\Omega$) describing the relative orientation of the
laboratory and intrinsic frame. The $\B$, $\G$ variables are given by mean
values of the operators $Q_0= \sum^A_{i=1} (3z^2_i-r_i^2)$
and $Q_2=\sum^A_{i=1} \sqrt{3}(x^2_i-y^2_i)$ as follows
\begin{eqnarray}
\label{eq:bet1} &\beta\cos\gamma=c q_0, \quad\quad q_0= \langle\Phi| Q_0 |\Phi\rangle\\
\label{eq:bet2} &\beta\sin\gamma=c q_2, \quad\quad q_2=\langle\Phi| Q_2 |\Phi\rangle
\end{eqnarray}
with a conventional factor  $c=\sqrt{\pi/5}/A\overline{r^2}$ 
where
$\overline{r^2}=3/5r^2_0A^{2/3}$, $r_0=1.2$~{\rm fm}.

One should keep in mind that  in some theoretical approaches, e.g. in the
geometrical collective (Frankfurt) model \cite{x85eis}, the deformation
variables $\B$, $\G$ do not have a direct relation to a nuclear shape or
mass distribution. Within a framework of the interacting boson model
\cite{x87iac} the $\B$, $\G$ variables introduced by means of the so called
coherent states are related rather with properties of valence nucleons and
not of a spatial distribution of a nuclear density.

The general properties of the quadrupole collective space as well as of functions and
operators depending on the quadrupole variables can be found e.g. in
\cite{2009PR08}. The most important, from the point of view of physical
applications, is a Hamiltonian which we call the General Bohr Hamiltonian
(GBH) and which can be expressed in the intrinsic frame as
\beqn\label{eq:gbh}
H_{\rm Bohr}=T_{\rm vib}+T_{\rm rot}+{ V}
\eeqn
\begin{eqnarray}
\nonumber T_{\rm vib}&=-{1\over{2\sqrt{wr}}}\left\{ {1\over \B^4}\bigg[ \dbe \bigg(
\B^4\rnw {\Bgg}\bigg)\dbe - \dbe \bigg(\B^3\rnw
{\Bbg}\bigg)\dga\bigg]+
\right. \\
&\label{eq:tvib} \left. + {1\over \B\Sg}\bigg[ -\dga \bigg( \rnw \Sg
{\Bbg}\bigg)\dbe +
{1\over\B}\dga \bigg(\rnw\Sg
{\Bbb}\bigg)\dga\bigg]  \right\}\\
\label{eq:trot} T_{\rm rot}&={1\over 2}\sum_{k=1}^{3} I^2_k(\Omega)/J_k; \ \ \  \
J_k=4{B_k(\B,\G)}
\B^2\sin^2(\G -2\pi k/3)\\[2mm]
&\label{eq:wwr} \ \mbox{where}\quad\quad w=B_{\B\B}B_{\G\G}-B^2_{\B\G}; \ \ r=B_xB_yB_z 
\end{eqnarray} 
The operators
$I_k(\Omega), k=1,2,3$ are components of the angular momentum in the
intrinsic frame. The Hamiltonian (\ref{eq:gbh}) contains seven functions
that depend on deformation variables: the potential energy $V$ and six functions
$B_{\B\B},B_{\B\G},B_{\G\G}, B_k$, called mass parameters or
inertial functions. 

One possible way to determine these seven functions consists in assuming for
them a 'reasonable' form with some free parameters which are determined
through comparison of calculated and experimental collective properties. I
use another approach which is based on the ATDHFB (Adiabatic Time Dependent
HFB) theory and which aims at calculating these functions starting from a
microscopic theory. In this approach one does not introduce any additional
free parameters and the prediction of collective properties is based solely
on the knowledge of effective nucleon-nucleon interactions.

\subsection{The ATDHFB mass parameters}

In the following discussion it is assumed that the time evolution of a system is
determined through a time dependence of several collective variables $q_j$
(not necessarily the quadrupole ones from eqs.
(\ref{eq:bet1}-\ref{eq:bet2})). The ATDHFB theory, based on an assumption of
low collective velocities, gives an expression $
\frac{1}{2}\textstyle\sum_{k,j}B_{kj}\dot{q}_k\dot{q}_j $ which is bilinear
in velocities and which defines a metric tensor in the collective space. In
the next step this expression is used to calculate the Laplace-Beltrami
operator which is taken (up to the $\hbar^2$ factor) as a kinetic energy
part of a collective Hamiltonian. Functions $B_{kj}$ (mass parameters)
depend on collective variables. More details on the ATDHFB theory and mass
parameters can be found in e.g. \cite{2009PR08,2011BA45} and papers cited
therein. Below I briefly sketch some steps and give some formulas
which are needed to calculate the general Bohr Hamiltonian starting from the
UNEDF0 energy functional.

The so called cranking approximation  ignores the
Thouless-Valatin terms so that the mass parameters can be conveniently expressed
through derivatives of a generalized density matrix $R(\we{q})$ 
corresponding to the HFB state $\Phi(\we{q})$. The derivative
$\displaystyle\partial_{ q_k}R$ in the quasi-particle basis (in the doubled
space) has the form
\beqn
\left(\frac{\partial}{\partial q_k}R\right)_{\rm quasipart}
=F_k=\bpmax
0 & f_k\\
\tilde{f}_k & 0
\epmax, \ \ \ \tilde{f}_k=-f^*_k
\eeqn
and the mass parameters read
\beqn
B_{kj}=
\frac{\hbar^2}{2}\sum_{\mu\nu}\frac{f_{j,\mu\nu}f^*_{k,\mu\nu}
+f^*_{j,\mu\nu}f_{k,\mu\nu}}{E_\mu+ E_\nu}
\eeqn
where $E_{\mu,\nu}$ are quasi-particle energies.
If the matrix $\displaystyle\partial_{ q_k}R$ is known in a fixed
single-particle basis the matrix $F_k$ can be calculated as
\beqn
F_k=\clB \left( \partial_{q_k} R\right)_{\rm fixed, doubled} \clB^+
\eeqn
where
$\clB$ is the Bogolyubov matrix for $R$
\beqn
\clB= \bpmax
U^+ & V^+\\
V^T & U^T
\epmax.
\eeqn
Sometimes the following
alternative expression for $f_k$ is useful
\beqn
f_{k,\mu\nu}=\bra {\Phi} \alpha_\nu\alpha_\mu\ket{\partial_{q_k}\Phi}
\eeqn
where $\alpha_{\mu,\nu}$ are quasi-particle annihilation operators.

In the case of quadrupole variables one obtains the deformation-dependent HFB
state by constrained HFB calculations
\beqn
\label{eq:cons}\delta \langle \Phi | H_{\rm micr}| \Phi\rangle=0 \ \ \ \mbox{with} \ \ \
\langle \Phi | Q_j | \Phi\rangle=q_j, \ j=0,2
\eeqn
Then, it is easier first to discuss the vibrational mass parameters $B_{q_iq_k},
k=0,2$
from which $B_{\B\B}, B_{\B\G}, B_{\G\G}$ in the formulas
(\ref{eq:tvib}-\ref{eq:wwr}) can be calculated by a simple change of variables.
The required derivatives $\partial_{q_k}R$ should be calculated by numerical
differentiation (see \cite{1999YU09, 2004PR01,2011BA45}) but most often one
resorts to another (so called perturbative) approximation which relates
the derivatives of the generalized density matrix to derivatives of the induced
one-body Hamiltonian \cite{2009PR08}. The constraints (eq \ref{eq:cons})
lead to the extra term $-\lambda_j Q_j$ in the induced one-body Hamiltonian
and one can easily calculate a derivative with respect to $\lambda_j$:
\beqn
(f_{\lambda_j})_{\mu\nu}=\frac{1}{E_\mu+E_\nu}(w_j)_{\mu\nu}, \ \
\eeqn
where
\beqn
\label{eq:www}
w_j=U^+(Q_j)_{\rm fixed}V^*-(U^+(Q_j)_{\rm fixed}V^*)^{T}
\eeqn
Then, the derivatives $\partial_{q_k}R$ are calculated using the relation
\beqn
\frac{\partial}{\partial {q_k}}R=
\sum_j\frac{\partial \lambda_j}{\partial {q_k}}
\frac{\partial }{\partial {\lambda_j}}
R
\eeqn
and finally the derivatives ${\partial \lambda_j}/{\partial {q_k}}$ are 
obtained by inverting the matrix ${\partial q_k}/{\partial
{\lambda_j}}$, which can  also be  expressed through $w_j$, eq (\ref{eq:www})
\beqn
\frac{\partial q_k}{\partial
{\lambda_j}}= 
\Relp \sum_{\mu\nu}\frac{(w_k^*)_{\mu\nu}(w_j)_{\mu\nu}
}{E_\mu+E_\nu} .
\eeqn

The moments of inertia are given by the  Inglis-Belyaev formula
\beqn \label{eq:momin}
J_k=\sum_{\mu,\nu}\frac{|[U^+(j_k)_{\rm fixed} V^* - (U^+(j_k)_{\rm fixed}
V^*)^T]_{\mu\nu}|^2}{E_{\mu}+E_{\nu}}
\eeqn
where $(j_k)_{\rm fixed}$ is a matrix of the microscopic total angular
momentum.

In the case of the BCS approximation when the canonical basis is used 
formulas (\ref{eq:www}, \ref{eq:momin})
can be transformed to a simpler and better known form 
but in the general HFB approach (\ref{eq:www}, \ref{eq:momin}) are more
useful.

\subsection{The UNEDF0 Energy functional}

To construct the mean-field configurations I used the UNEDF0
energy density functional, which is one of the results of a large scale project UNEDF
\cite{xx13BO01}. The functional is described in detail in \cite{2010KO29}
and here I will elaborate only on some of its distinctive features. In the particle-hole
channel UNEDF0 is a 'standard' Skyrme-type functional \cite{2003BE12} with
the spin-orbit term treated as in the SkI parametrization \cite{1995RE03}.
The pairing interaction is modelled as a sum of the standard (volume) plus
density-dependent surface peaked
$\delta$ interaction
\begin{equation}
\frac{{V^{q}_0}}{2}\left[1+\bigg(1-\frac{\rho(\we{r})}{\rho_0}\bigg)\right]\delta(\we{r}_1-\we{r}_2),
\ \ q={\rm n}, {\rm p}; \ \ \rho_0=0.16~{\rm fm}^{-3}
\end{equation}
and the Lipkin-Nogami method is used to avoid the pairing 'collapse' for
magic nuclei and their neigbours.
The pairing strengths for protons and neutrons, $V^{\rm n,p}_0$ are fitted
simultaneously with other parameters determining the functional. A
truncation of the quasi-particle space, required due to a zero-range of the
pairing interaction is fixed by the condition for quasi-particle
energies $E_{\mu}< 60~{\rm MeV}$. Because it is well known that the ATDHFB mass
parameters are quite sensitive to a diffuseness of the occupation number
distribution I shall now present more details on the treatment of the pairing part of UNEDF0.

\begin{figure}[htp]
\includegraphics[scale=0.4]{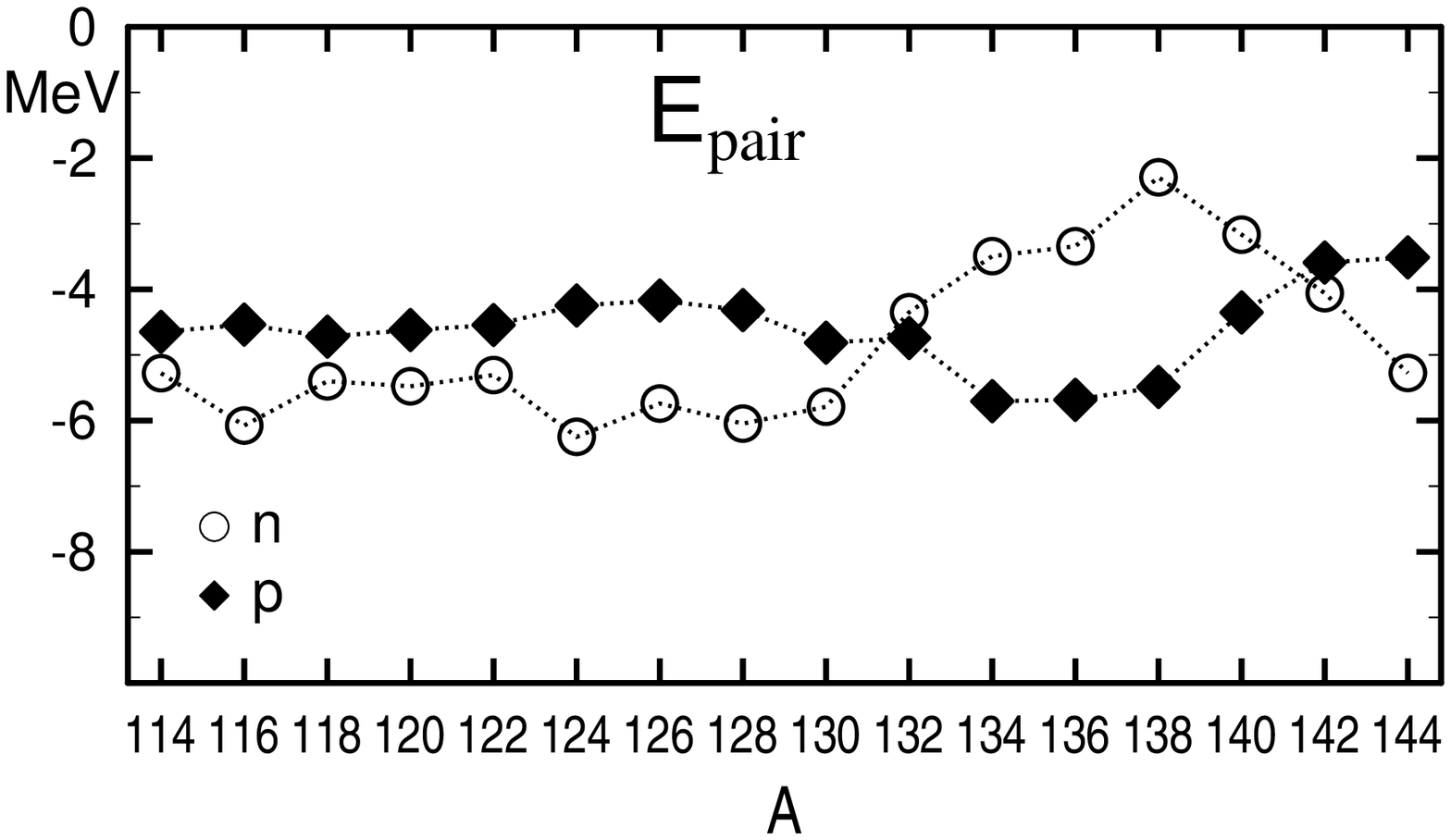}

\includegraphics[scale=0.4]{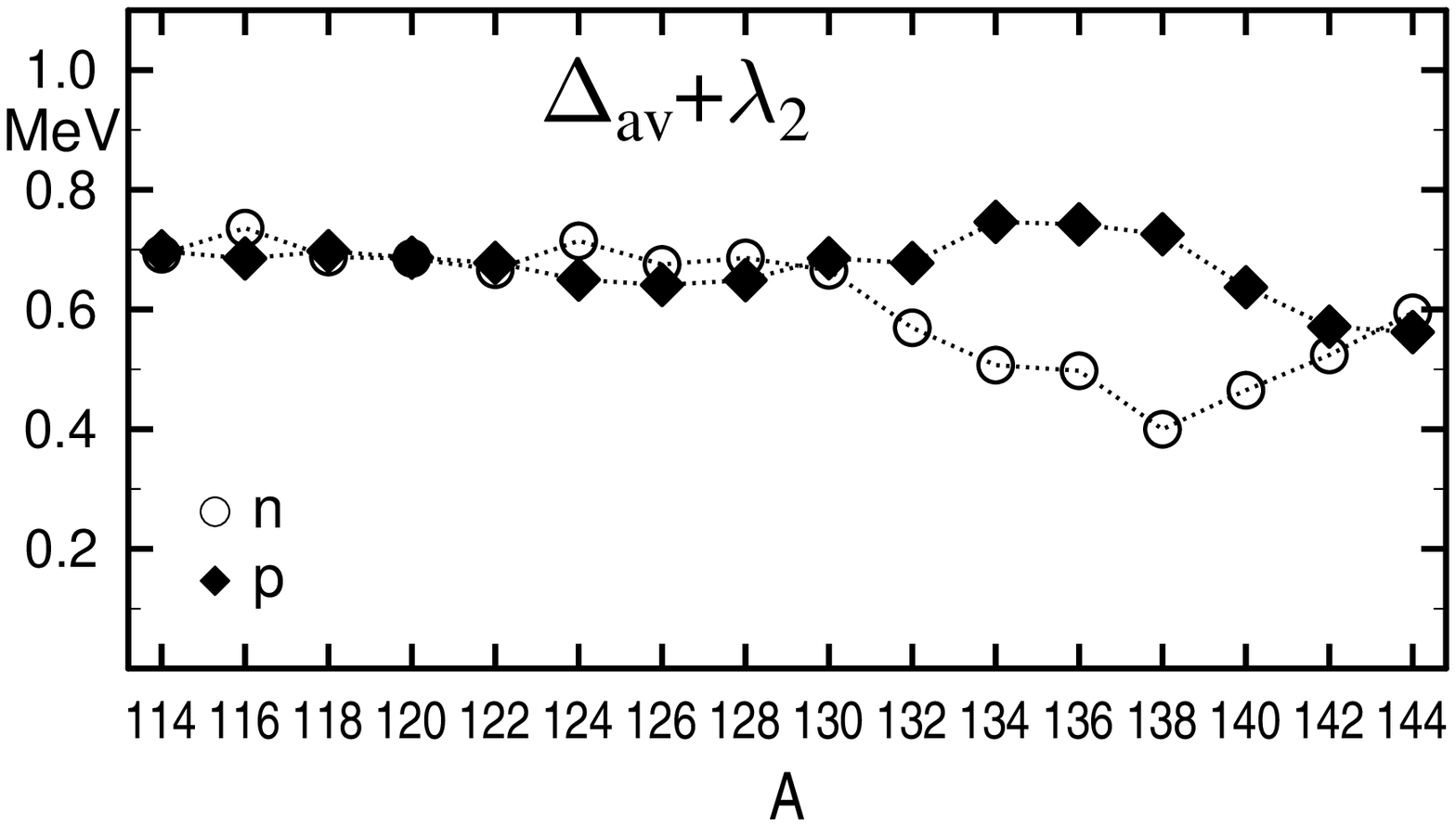}
\caption{\label{fig:pairprop}Plot of the pairing energy (upper panel) and of
the effective pairing gap $\Delta_{\rm av}+\lambda_{2,LN}$ calculated for protons (\fulldiamond)and neutrons
(\opencircle) at the
deformation corresponding to a minimum of the potential energy.}
\end{figure}

The binding energies of the considered Xe isotopes are reproduced quite well by
the UNEDF0 functional. The RMSD (root mean square deviation) for 16 nuclei
is equal to 0.454~MeV with the largest error $B_{\rm th}-B_{\rm
exp}=-0.71$~MeV for the \nuc{134}{Xe} isotope. The chain of isotopes
contains \nuc{136}{Xe} with a magic number $N=82$ of neutrons but it appears
that due to the Lipkin-Nogami (LN) prescription the changes of the pairing
properties along the chain are quite smooth. This can be seen in figure
\ref{fig:pairprop} where I plot the neutron and proton pairing energy vs
the mass number. In addition I show a plot of the quantity $\Delta_{\rm
av}+\lambda_{2,LN}$, where $\Delta_{\rm av}=\sum
u_{\mu}v_{\mu}\Delta_{\mu}/\sum u_{\mu}v_{\mu}$ and $\lambda_{2,LN}$ is a
coefficient determined in the LN method. This quantity can be treated as an
estimation of the pairing gap within the LN method, for more details see
\cite{2000BE32}.

In conclusion I want to mention two newer functionals UNEDF1 \cite{2012KO06}
and UNEDF2 \cite{2012KO06, 2014KO13} which were constructed by extending the
empirical dataset used in the fitting procedure. In the case of the UNEDF1
functional new data on a few fission isomers was added while for the UNEDF2
several single-particle level splittings were additionally considered.
However, the RMSD for binding energies is significantly lower (around 1.4
MeV) for UNEDF0 than for UNEDF1 and UNEDF2 (around 1.9 MeV) hence the UNEDF0
functional seems to be a good choice for a pilot study of collective
properties in the region of medium-heavy nuclei.
A further detailed study on the consequences of UNEDF1 and UNEDF2 for
collective nuclear properties is currently in progress.

\section{\label{sec:res}Results of calculations, comparison with experiment}

The values of inertial functions and potential energy which enter the
General Bohr Hamiltonian were calculated at 144
points forming a regular grid in the sextant $(0\le \beta\le 0.65)\times
(0\le \gamma \le 60^{\circ})$ in the deformation plane. The distance between
the points is 0.05 and $6^{\circ}$ in the $\beta$ and $\gamma$ directions,
respectively. The mean-field wave functions were obtained using the code
HFODD ver. 2.49t, see \cite{xx12SC01} and references therein.

\subsection{Potential energy surfaces}

\begin{figure*}[htp]
(a) \parbox[c]{1.0\textwidth}{
\includegraphics[scale=0.4]{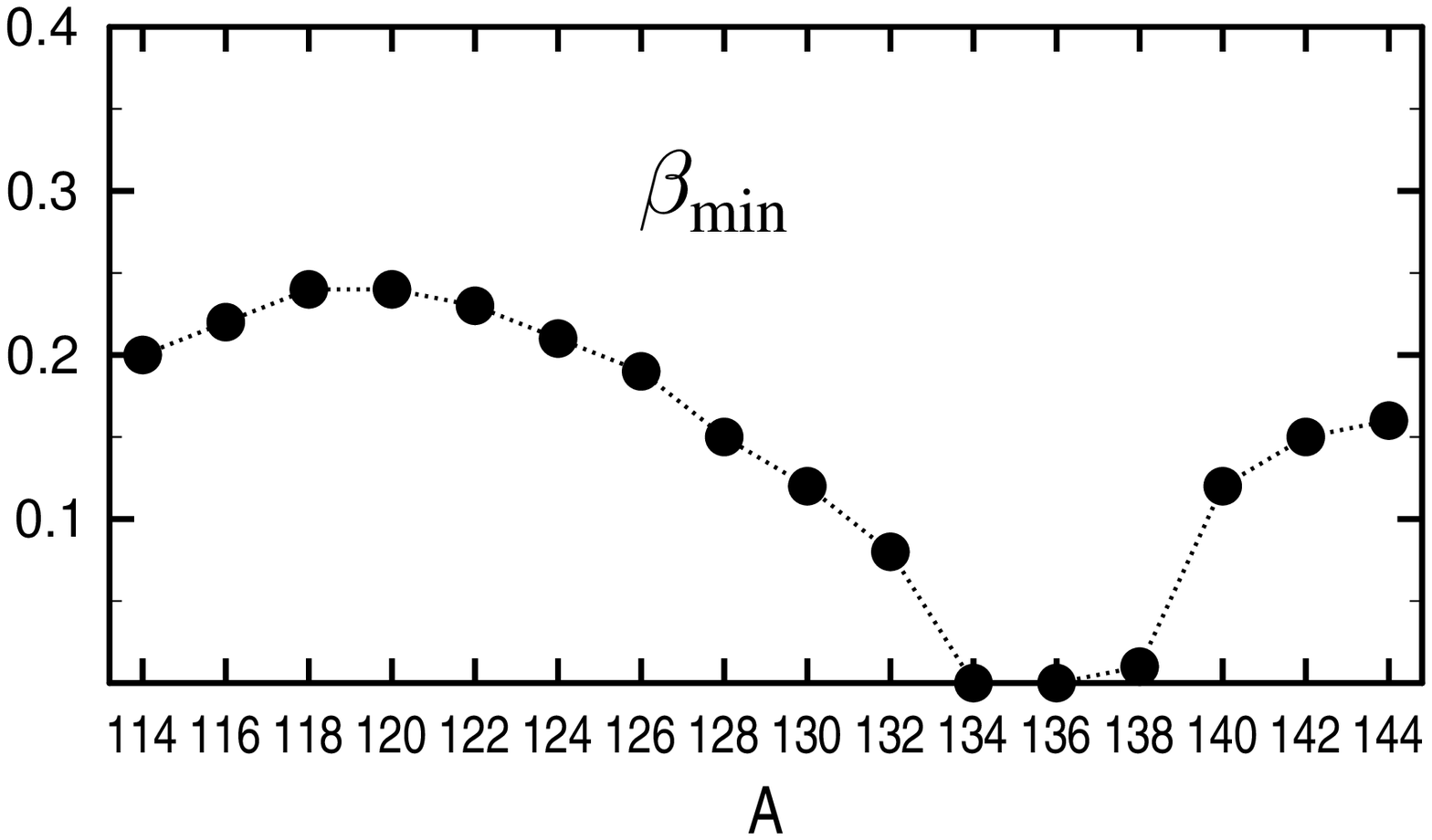}
\includegraphics[scale=0.4]{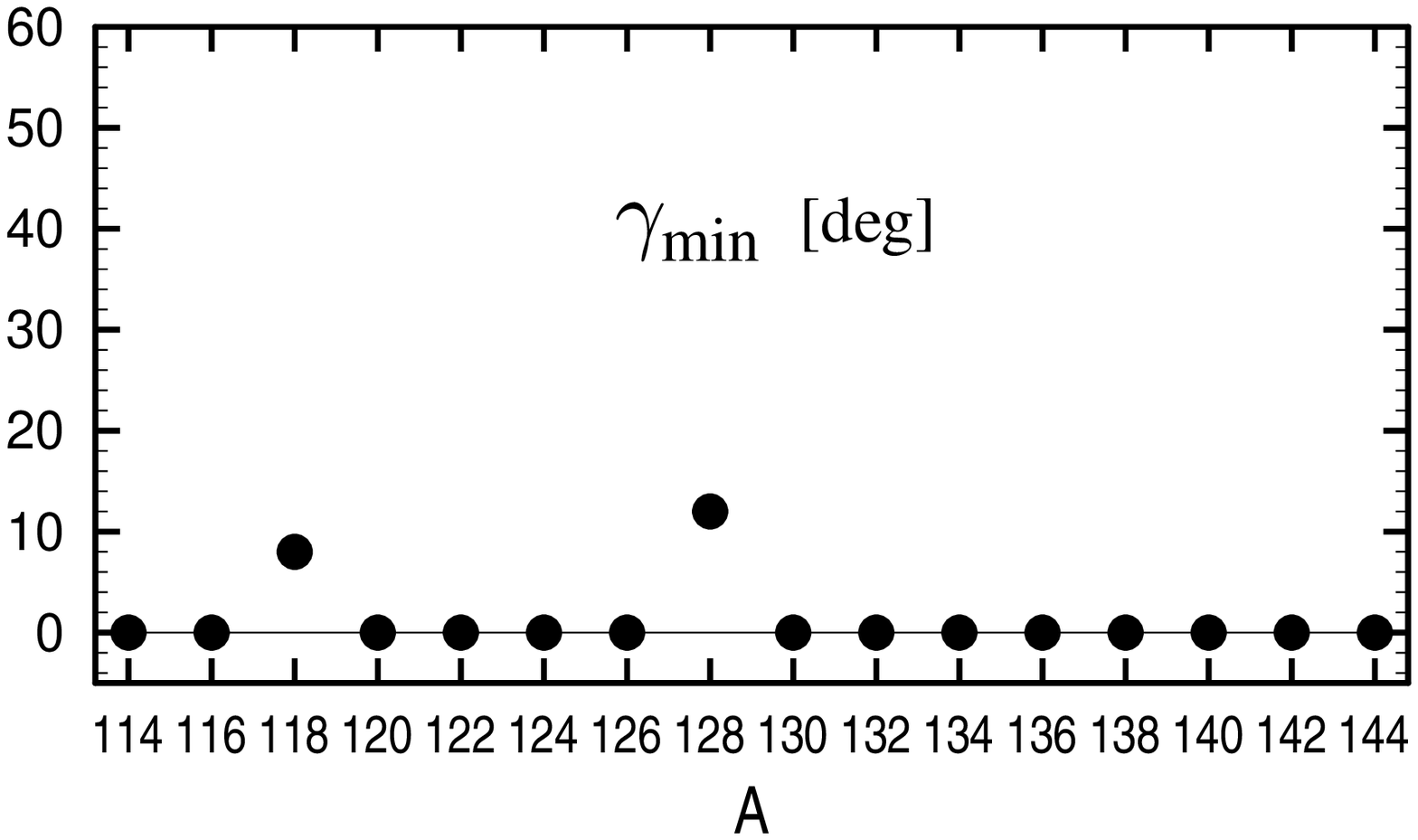}}\vspace{-4mm}

(b)\parbox[c]{0.7\textwidth}{\includegraphics[scale=0.4]{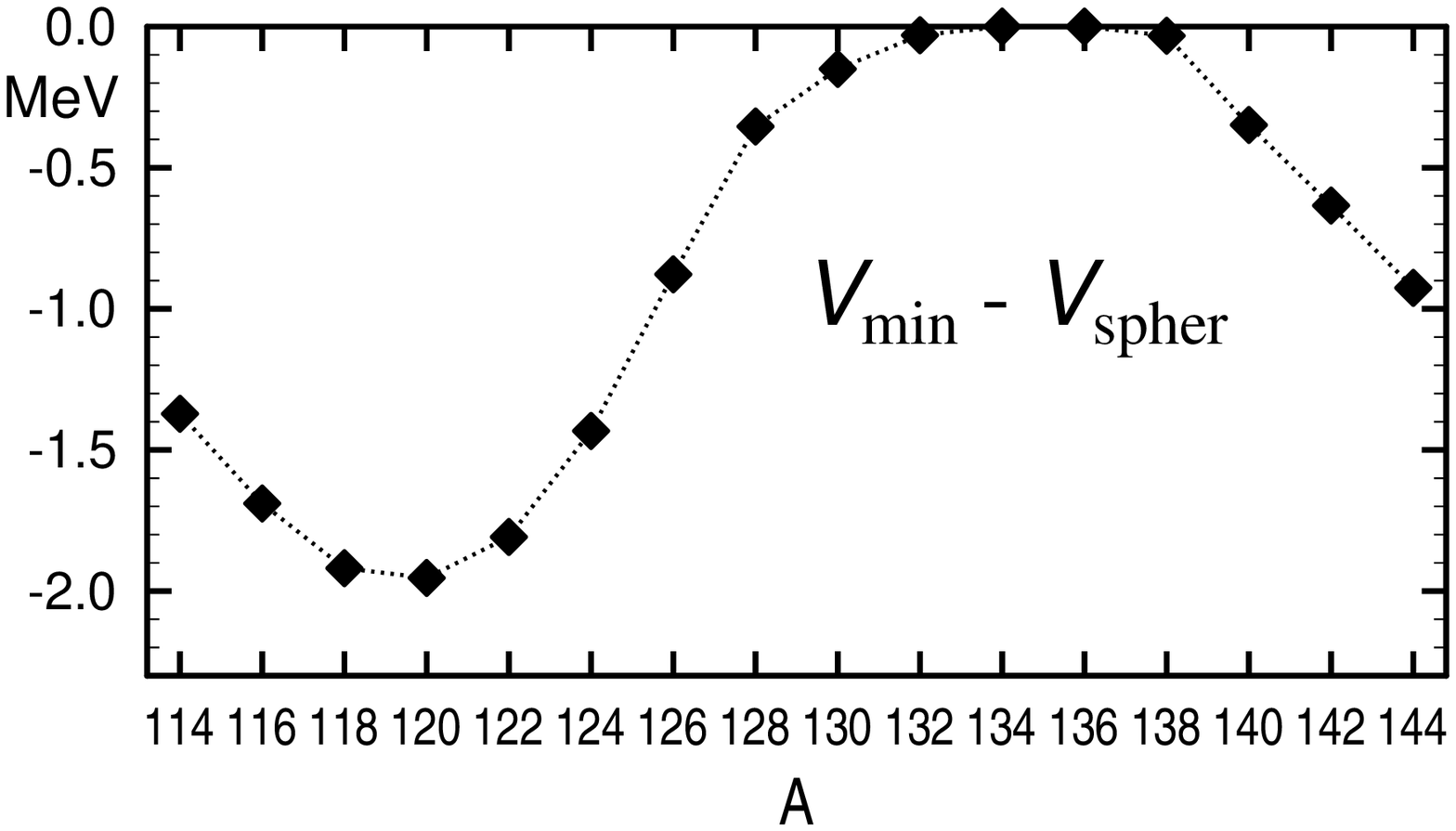}}
\caption{\label{fig:xe-miny}a) Value of the deformation $\beta_{\rm min}$
(left panel) and $\gamma_{\rm min}$ (right panel) at
the minimum of the potential energy. 
(b) Depth (relative to a value at a spherical shape) of minima of the
potential energy for the \nuc{114-144}{Xe} nuclei.}
\end{figure*}

\begin{figure*}[htp]
\includegraphics[scale=0.6]{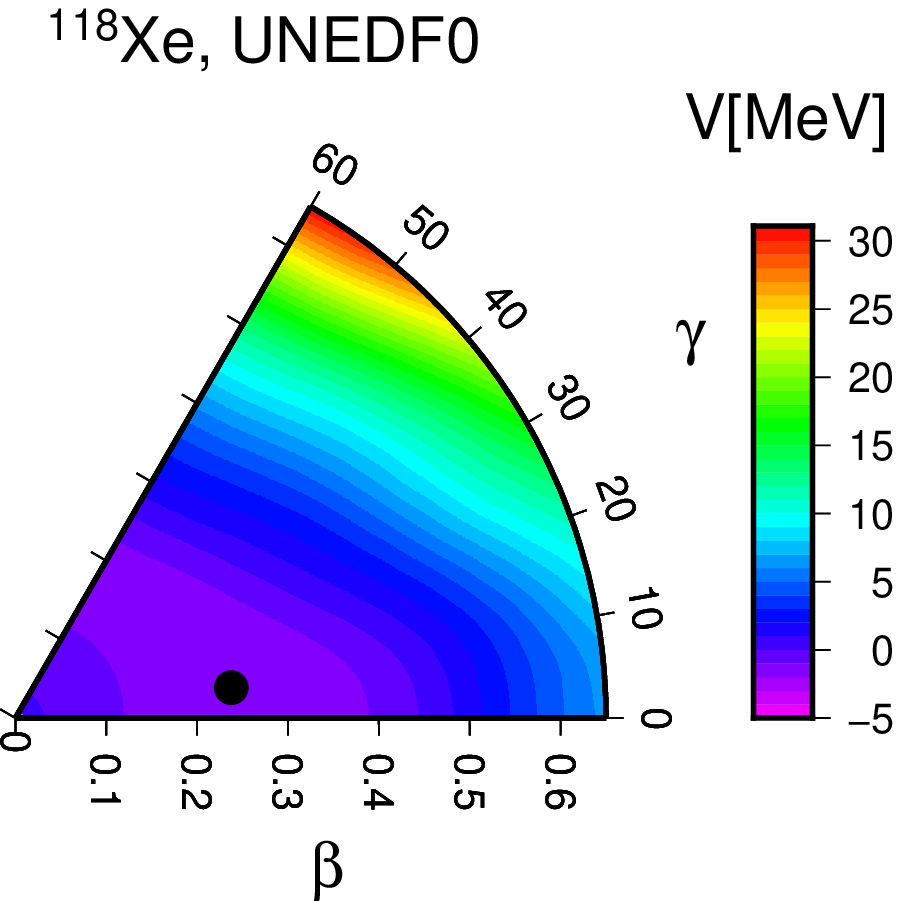}\hspace{2cm}
\includegraphics[scale=0.6]{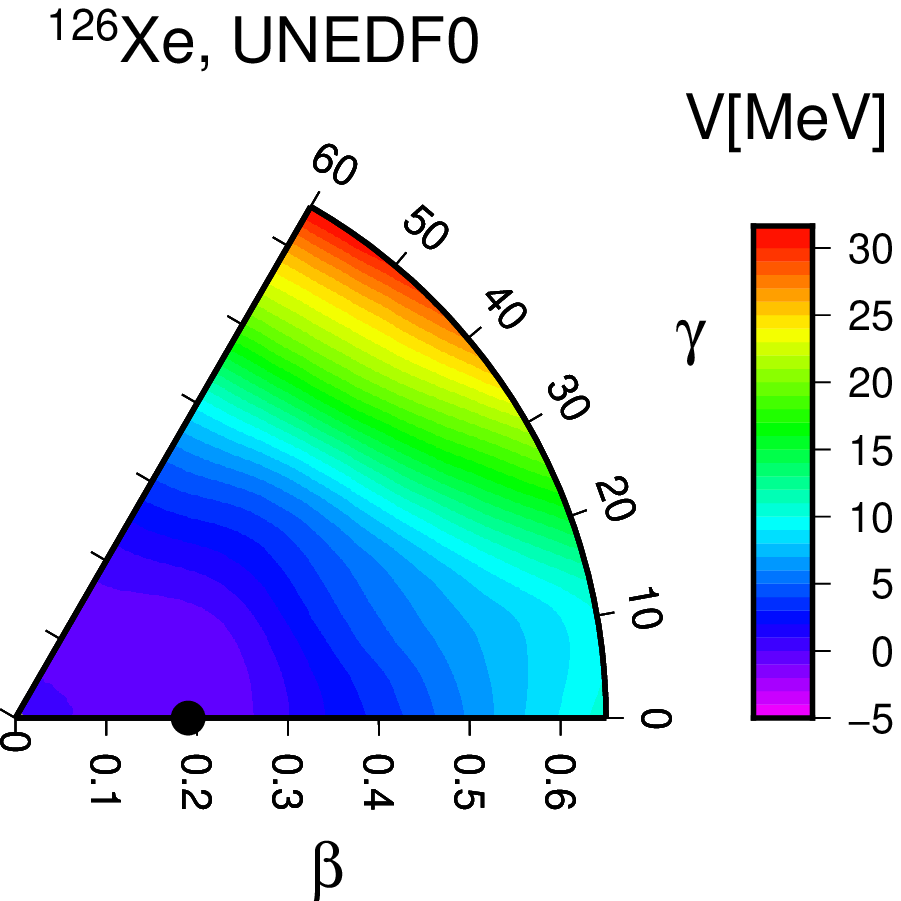}

\caption{\label{fig:potf1}Plot of the potential energy (relative to a spherical
shape value) for the \nuc{118}{Xe} and \nuc{126}{Xe} isotopes.}
\end{figure*}

\begin{figure*}[htp]
\includegraphics[scale=0.6]{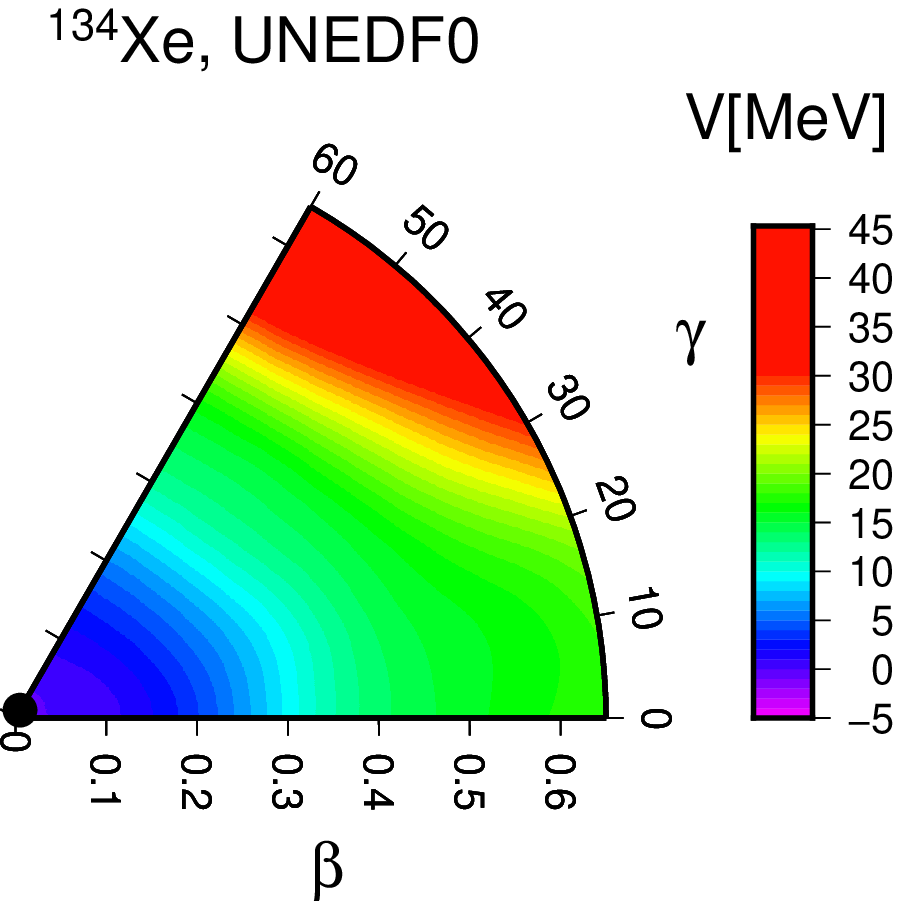}\hspace{2cm}
\includegraphics[scale=0.6]{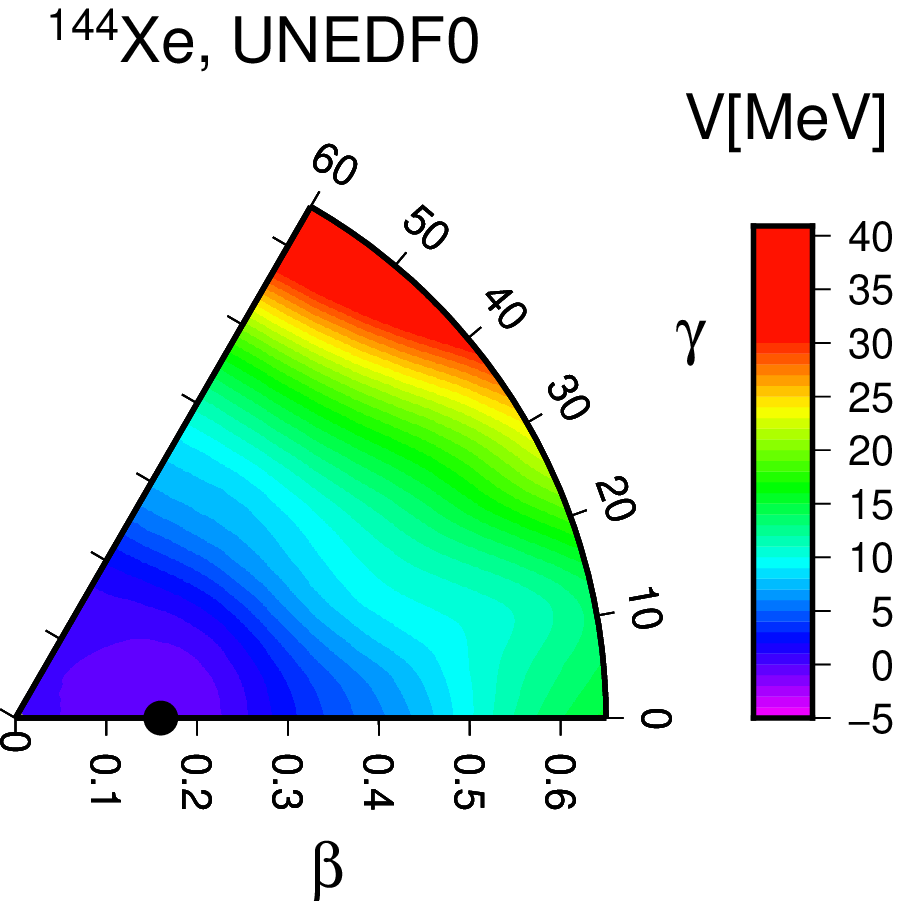}
\caption{\label{fig:potf2}Plot of the potential energy (relative to a spherical
shape value) for the \nuc{134}{Xe} and \nuc{144}{Xe} isotopes.}
\end{figure*}

As can be seen in \fref{fig:xe-miny} there are three nuclei
\nuc{134-138}{Xe} with a spherical minimum of the potential energy. Others
exhibit deformed minima with $\B_{\rm min}$ in the range $0.1-0.25$, mostly
on the prolate axis except for \nuc{118}{Xe} and \nuc{128}{Xe} which have
slightly nonaxial minima with $\G_{\rm
min}=8^{\circ}$ and $12^{\circ}$, respectively. The depths of the minima
(relative to a spherical shape) are less than $2$~MeV. In
figures~\ref{fig:potf1} and \ref{fig:potf2} I show full plots of the
potential energy on the deformation space for a representative sample of
four isotopes. One can notice a rather weak dependence of the potential energy
on the $\G$ variable ($\G$ softness), especially for lighter isotopes.

\subsection{Collective energy levels}

Having calculated the potential energy and mass parameters I performed a
numerical diagonalization of the resulted Bohr Hamiltonian using the method
described in \cite{1999PR03,2004PR01}. The obtained eigenvalues can be
directly compared with excited energy levels of positive parity and the
corresponding collective wave functions can be then used to calculate matrix
elements of various operators, in particular of the operators of
electromagnetic E2 transitions.

\begin{figure}[htp]
\includegraphics[scale=0.6]{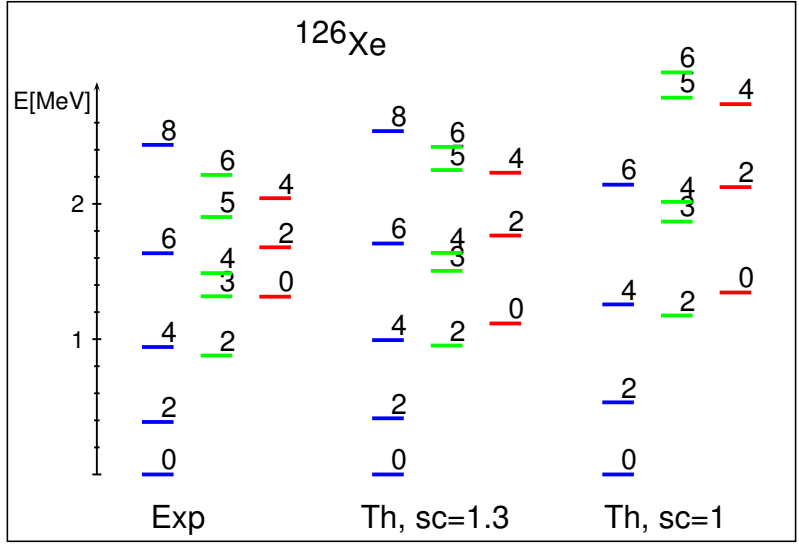}
\caption{\label{fig:drab126}Energy levels in \nuc{126}{Xe}. Comparison of 
theoretical results obtained using the
scaling of mass parameters ($sc=1.3$), without such scaling ($sc=1$) and
experimental data.}
\end{figure}

\begin{figure}[htp]
\includegraphics[scale=0.6]{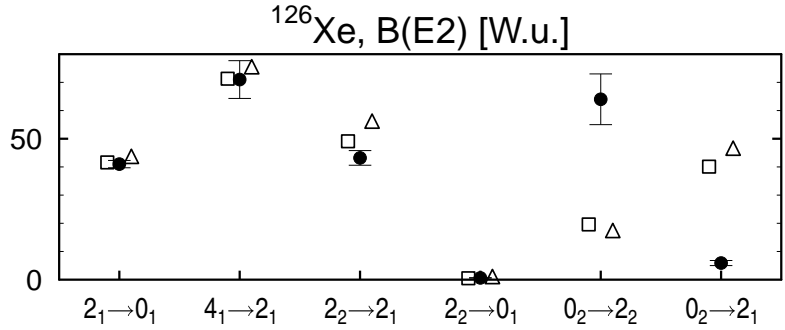}
\caption{\label{fig:be2-126}Selected E2 transitions in the \nuc{126}{Xe}
nucleus. Theoretical results with the scaling factor ($sc=1.3$),  without
one
($sc=1$) and experimental data \cite{2011CO07} are shown with
(\opensquare),(\opentriangle) and (\fullcircle), respectively}
\end{figure}

I should add that all mass parameters
(vibrational and rotational) were multiplied before the diagonalization of the
Bohr Hamiltonian by
a constant factor $1.3$. The commonly quoted reasons for introducing such a
factor refer to a simulation of the effects of including the Thouless-Valatin
terms in the ATDHFB method or/and the effects of the so called pairing
vibrations, see e.g. \cite{1999LI38,2004PR01,2009PR08,2007PR07,2010LI09}.
Some rather crude estimations of these effects give the value of the factor in
the range $1.2-1.5$.
However, due to a lack of sufficiently quantitative calculations
this factor must be treated as an additional parameter of the theory.
Before presenting the results for the whole chain of Xe isotopes I will show 
consequences of introducing the scaling factor for energy spectra and B(E2)
probabilities in the case of \nuc{126}{Xe}.
\Fref{fig:drab126} contains plots of bands built on $0_1$, $2_2$ and
$0_2$ levels. There are two sets of theoretical results: obtained with the
scaling factor ($sc=1.3$) and without the scaling ($sc=1$). One can see that
the scaling produces a 'shrinking' of the spectra leaving a general picture
similar in both cases. In addition, one can see that the scaling leads to a better
agreement with experimental data (showed in \fref{fig:drab126} as well).
A sample of $B(E2)$ results (theoretical  with and without scaling and
experimental) is shown in \fref{fig:be2-126}. The sample contains 
cases with both good and worse agreement between theory and experiment. It
can be seen that the effect of the scaling  the mass parameters is much smaller
on the B(E2) probabilities than on the values of level energies.

Then I compare theoretical energies of several low lying low spin levels
($2_1, 4_1, 2_2, 0_2, 0_3$) with experimental data \cite{nndc0914} for the
considered chain of Xe isotopes.
These levels were chosen because of their role in analysing the band
structure of nuclear spectra.

\begin{figure}[htp]
\includegraphics[scale=\scala]{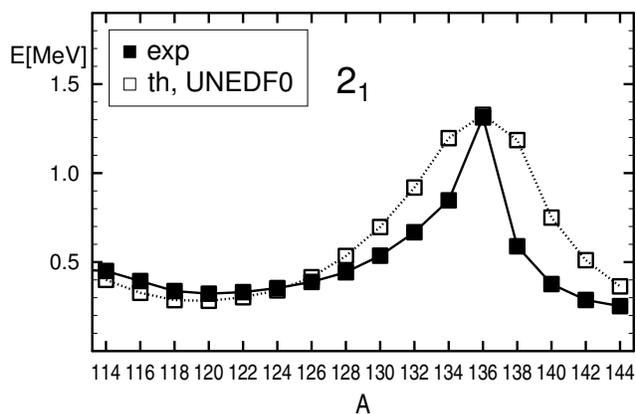}
\caption{\label{fig:lev21}Experimental (\fullsquare) \cite{nndc0914} and
theoretical (\opensquare)  energy of the $2^{+}_1$ level in the \nuc{114-144}{Xe} nuclei.}
\end{figure}

\begin{figure}[htp]
\includegraphics[scale=\scala]{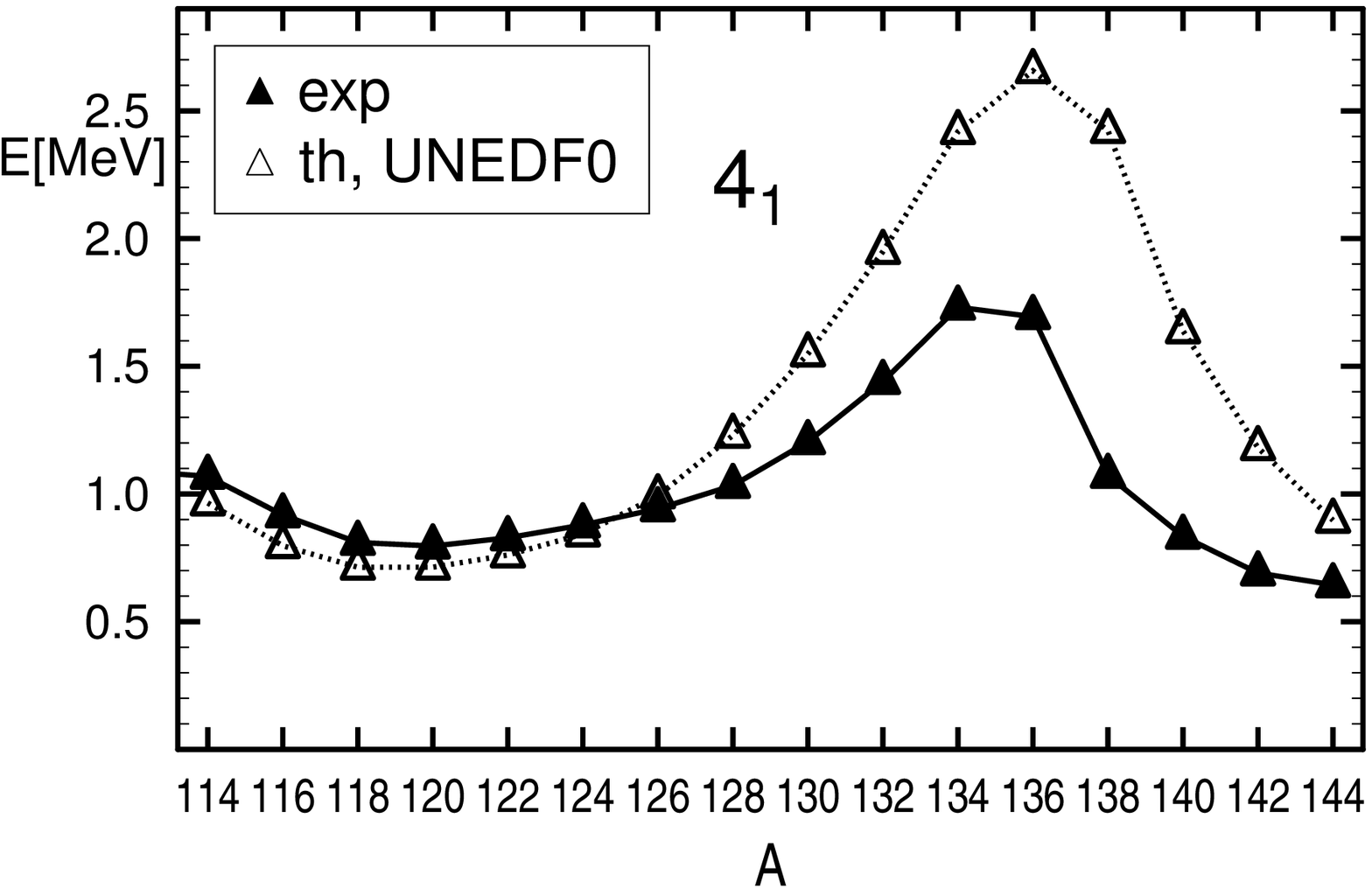}
\caption{\label{fig:41}Same as in \fref{fig:lev21}, but for the $4^{+}_1$
level.}
\end{figure}

\begin{figure}[htp]
\includegraphics[scale=\scala]{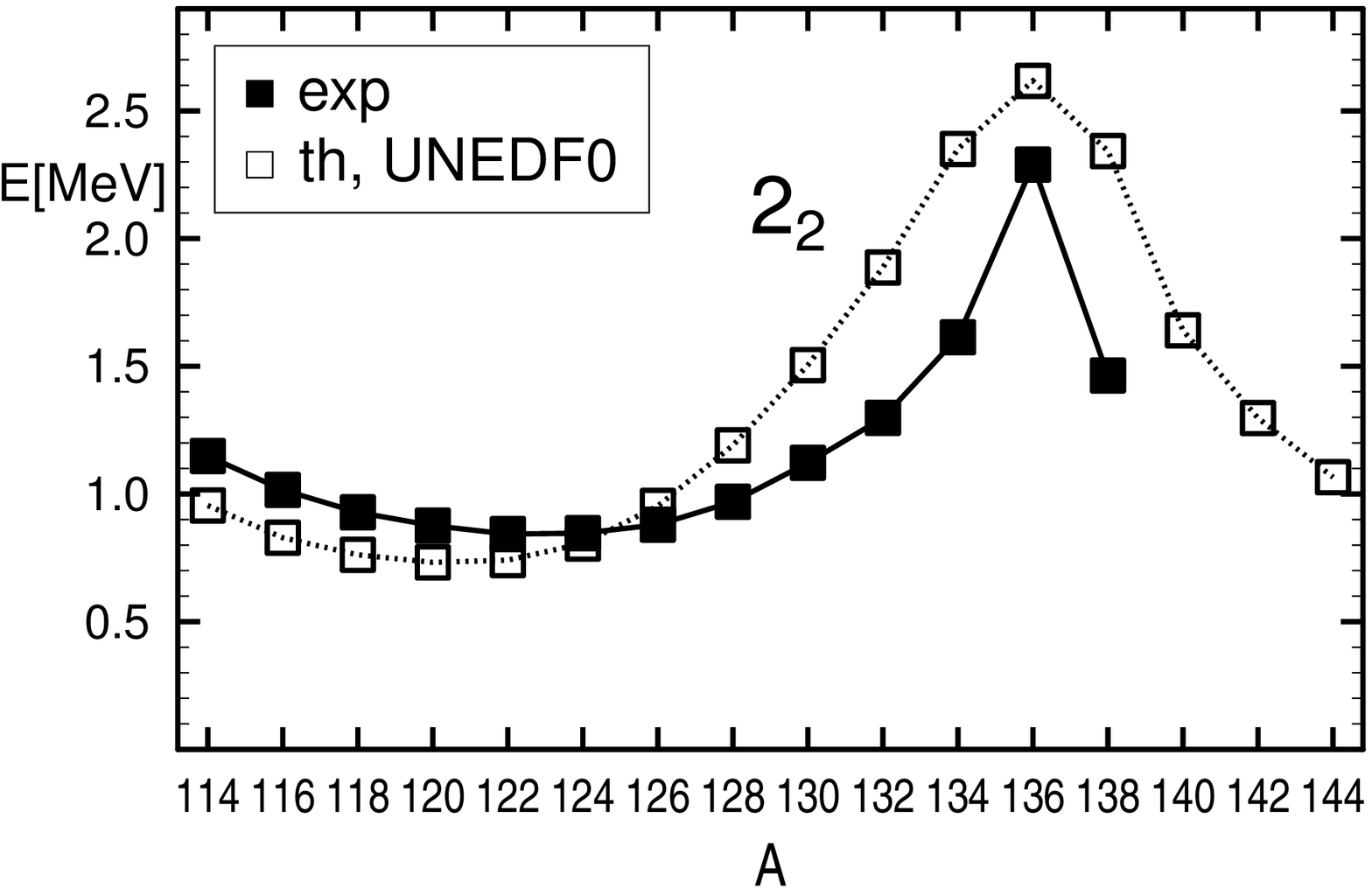}
\caption{\label{fig:lev22}Same as in \fref{fig:lev21}, but for the $2^{+}_2$
level.}
\end{figure}

\begin{figure}[htp]
\includegraphics[scale=\scala]{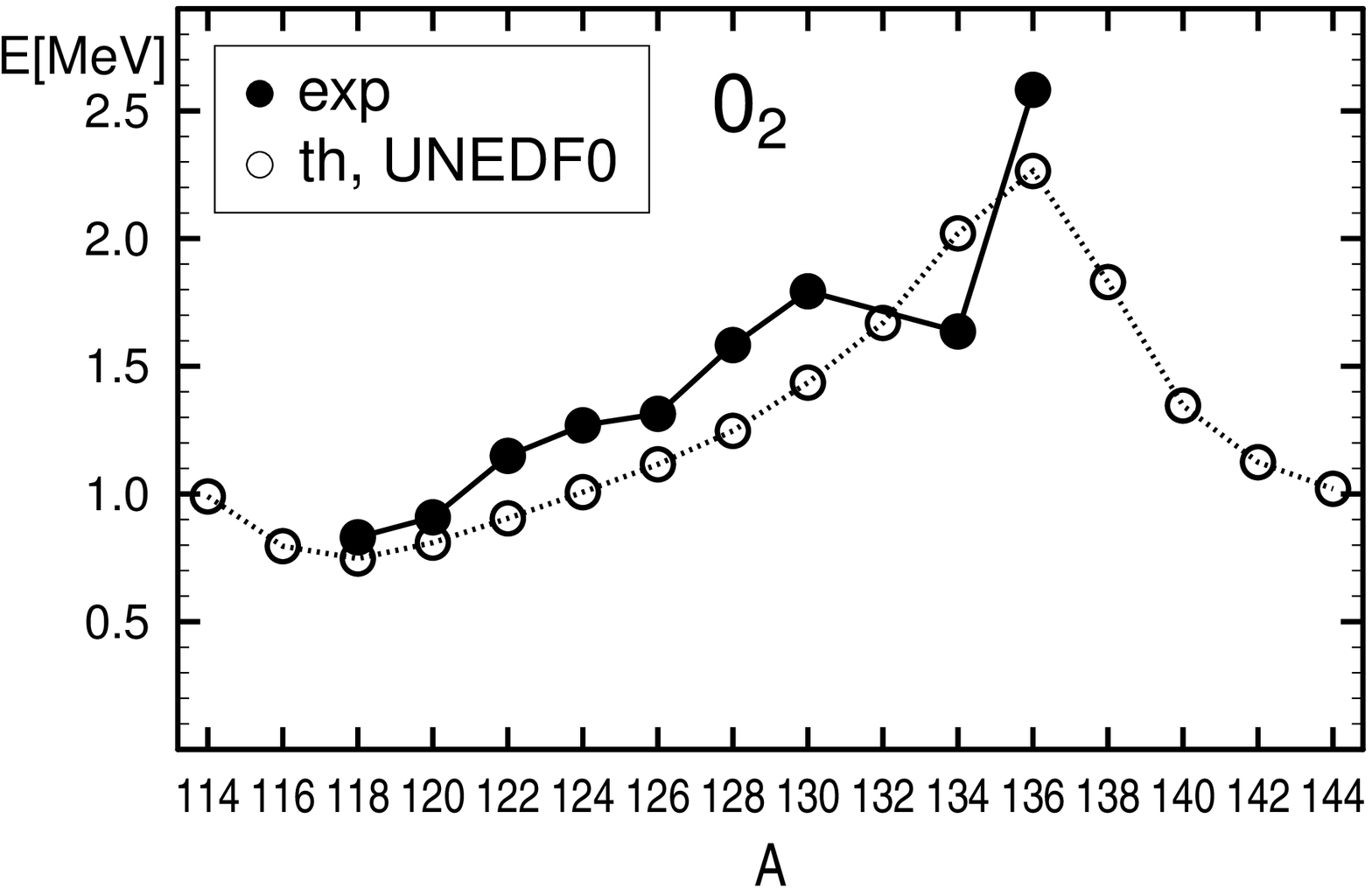}
\caption{\label{fig:02}Same as in \fref{fig:lev21}, but for the $0^{+}_2$
level.}
\end{figure}

\begin{figure}[htp]
\includegraphics[scale=\scala]{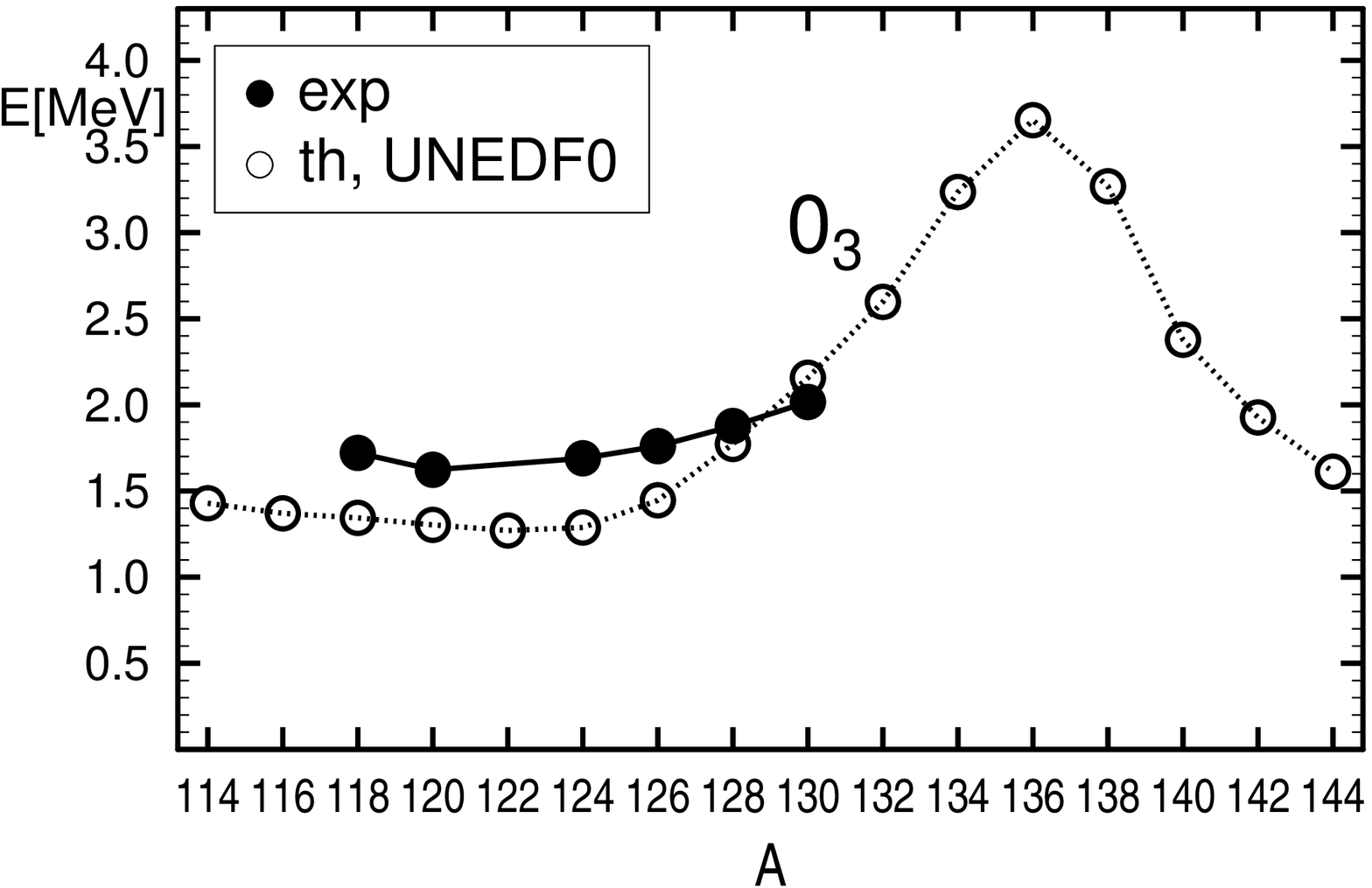}
\caption{\label{fig:03}Same as in \fref{fig:lev21}, but for the $0^{+}_3$
level.}
\end{figure}

One can conclude from plots in figures \ref{fig:lev21}-\ref{fig:03} that in
general the theoretical results are in good agreement with experimental data, especially for the lighter part of the isotope chain (up to $A=130$).
One should also keep in mind that I do not fit any parameters to
collective properties.  
Some significant discrepancies in the vicinity of $N=82$ number of neutrons
are not unexpected because the ATDHFB theory tends to perform better for
more collective nuclei, i.e. with a larger number of valence nucleons (let
us recall that there are only four valence protons in Xe isotopes). This
effect is connected with an assumption of adiabatic motion of all nucleons
in the varying mean-field. This assumption is strongly affected by a
presence of closed shells.

\subsection{E2 transitions}

\begin{figure}[htp]
\includegraphics[scale=\scalb]{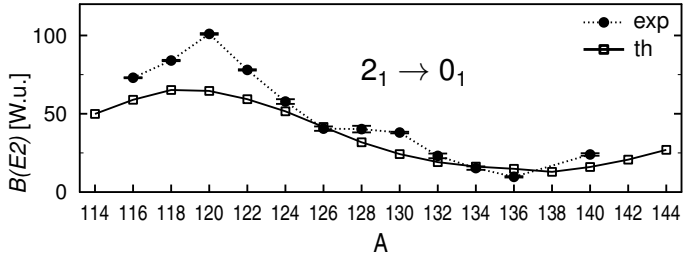}
\caption{\label{fig:e2fi}Experimental (\fullcircle) \cite{nndc0914} and
theoretical (\opensquare)  B(E2) probability (in W.u.) for the
transition $2^{+}_1 \rightarrow 0^{+}_1$ level in the \nuc{114-144}{Xe} nuclei.}
\end{figure}

\begin{figure}[htp]
\includegraphics[scale=\scalb]{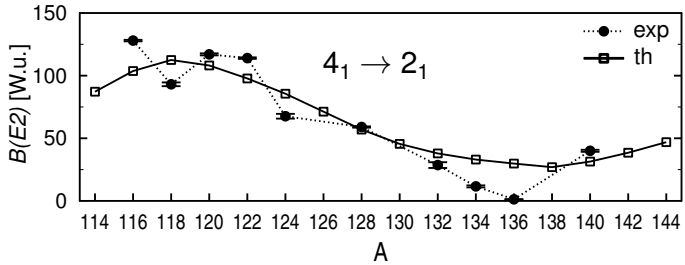}
\caption{\label{fig:e2se}Experimental (\fullcircle) \cite{nndc0914} and
theoretical (\opensquare)  B(E2) probability (in W.u.) for the
transition $4^{+}_1 \rightarrow 2^{+}_1$ level in the \nuc{114-144}{Xe} nuclei.}
\end{figure}

A detailed analysis of experimental data on electromagnetic transitions can
provide important information about excited levels, see e.g.
\cite{2012WR03}. In the case of quadrupole excitations the most important
are E2 transitions which are described by the collective operator
\begin{equation}
Q^{\rm (charge)}_{2\mu}(\B,\G)=
\langle \Phi(\B,\G)|e\textstyle \sum_{i=1}^{Z}r_i^2
Y_{2\mu}(\theta_i,\phi_i)|\Phi(\B,\G)\rangle
\end{equation}
I present the results of calculations of the B(E2) reduced transition
probabilities for transitions $2_1\rightarrow 0_1$ and $4_1\rightarrow 2_1$
as well as their comparison with evaluated experimental data from \cite{nndc0914} in
figures \ref{fig:e2fi} and \ref{fig:e2se}. Again one can see that
theoretical calculations reproduce the general behaviour of B(E2) quite well even as no free parameters (e.g. effective charges) were used.

In the case of some  Xe isotopes there are much more extensive experimental
data on E2 transitions, see e.g. for \nuc{126}{Xe} \cite{2011CO07} and for
\nuc{128}{Xe} \cite{1993SR01,2009CO24} but I postpone a discussion of them to
a subsequent publication.

\section{Conclusions}

The paper presents the results of the first attempt to apply the UNEDF0 energy
functional to the theory of a nuclear collective motion. A correct description
of the properties of the long chain of  Xe isotopes considered is a demanding
challenge to the theory, in particular to a framework with no free parameters that
could be fitted to experimental data on collective levels. It can be argued that the
results  shown in \sref{sec:res} are quite satisfactory and reproduce well
the general tendencies seen in the energy spectra as well as E2 transitions in the Xe
isotopes, with some exceptions around the semi-magic \nuc{136}{Xe} isotope.
This contribution contains only a part of the obtained theoretical results, a more
detailed analysis of the \nuc{120-128}{Xe} nuclei is in preparation.

\ack

The work was supported in part by the Narodowe Centrum Nauki (Polish National Center for
Scientific Research), grant No UMO-2013/10/M/ST2/00427. The author is
grateful to colleagues from the Department of Theoretical Physics UMCS Lublin
and from  NCBJ Warszawa-\'Swierk for an access to their computing facilities and
to Jacek Pr\'ochniak his for careful reading of the manuscript.

\section*{References}

\providecommand{\newblock}{}


\begin{thebibliography}{10}
\expandafter\ifx\csname url\endcsname\relax
  \def\url#1{{\tt #1}}\fi
\expandafter\ifx\csname urlprefix\endcsname\relax\def\urlprefix{URL }\fi
\providecommand{\eprint}[2][]{\url{#2}}

\bibitem{2010KO29}
Kortelainen M, Lesinski T, More J, Nazarewicz W, Sarich J, Schunck N, Stoitsov
  M~V and Wild S 2010 {\em Phys. Rev. C\/} {\bf 82} 024313

\bibitem{2004PR01}
Pr\'{o}chniak L, Quentin P, Samsoen D and Libert J 2004 {\em Nucl. Phys. A\/} {\bf
  730} 59

\bibitem{2009PR08}
Pr\'{o}chniak L and Rohozi\'{n}ski S~G 2009 {\em J. Phys. G (London)\/} {\bf 36} 123101

\bibitem{2010RO20}
Rodriguez T~R and Egido J~L 2010 {\em Phys. Rev. C\/} {\bf 81} 064323

\bibitem{2013YA05}
Yao J~M, Bender M and Heenen P~H 2013 {\em Phys. Rev. C\/} {\bf 87} 034322

\bibitem{2008BE29}
Bender M and Heenen P~H 2008 {\em Phys. Rev. C\/} {\bf 78} 024309

\bibitem{2014PA10}
Papakonstantinou P, Hergert H, Ponomarev V~Y and Roth R 2014 {\em Phys. Rev.
  C\/} {\bf 89} 034306

\bibitem{2007PA08}
Papakonstantinou P, Roth R and Paar N 2007 {\em Phys. Rev. C\/} {\bf 75} 014310

\bibitem{1977RO30}
Rohozi\'{n}ski S~G, Dobaczewski J, Nerlo-Pomorska B, Pomorski K and Srebrny J 1977
  {\em Nucl. Phys. A\/} {\bf 292} 66

\bibitem{1999PR03}
Pr\'{o}chniak L, Zajac K, Pomorski K, Rohozi\'{n}ski S~G and Srebrny J 1999 {\em Nucl.
  Phys. A\/} {\bf 648} 181

\bibitem{2012HI02}
Hinohara N, Li Z~P, Nakatsukasa T, Niksic T and Vretenar D 2012 {\em Phys. Rev.
  C\/} {\bf 85} 024323

\bibitem{2013BO24}
Bonatsos D, Georgoudis P~E, Minkov N, Petrellis D and Quesne C 2013 {\em Phys.
  Rev. C\/} {\bf 88} 034316

\bibitem{1988VO09}
von Brentano P, Gelberg A, Harissopulos S and Casten R~F 1988 {\em Phys. Rev.
  C\/} {\bf 38} 2386

\bibitem{1991KR10}
Krips W, Frank W, Lieberz W and von Brentano P 1991 {\em Nucl. Phys. A\/} {\bf
  529} 485

\bibitem{2008SO12}
Sorgunlu B and Isacker P~V 2008 {\em Nucl. Phys. A\/} {\bf 808} 27

\bibitem{2011CO07}
Coquard L, Rainovski G, Pietralla N, Ahn T, Bettermann L, Carpenter M~P,
  Janssens R~V~F, Leske J, Lister C~J, Moller O, Moller T, Rother W, Werner V
  and Zhu S 2011 {\em Phys. Rev. C\/} {\bf 83} 044318

\bibitem{1996PA03}
Pan X~W, Ping J~L, Feng D~H, Chen J~Q, Wu C~L and Guidry M~W 1996 {\em Phys.
  Rev. C\/} {\bf 53} 715

\bibitem{2008ME04}
Meng X, Wang F, Luo Y, Pan F and Draayer J~P 2008 {\em Phys. Rev. C\/} {\bf 77}
  047304

\bibitem{2011HI05}
Higashiyama K and Yoshinaga N 2011 {\em Phys. Rev. C\/} {\bf 83} 034321

\bibitem{x52bo01}
Bohr A 1952 {\em Mat. Fys. Medd. Dan. Vid. Selsk.\/} {\bf 26} no.~14, 1

\bibitem{x69bor}
Bohr A and Mottelson B~R 1969 {\em Nuclear Structure, vol.~I\/} (Benjamin Inc)

\bibitem{1976KA30}
Kaniowska T, Sobiczewski A, Pomorski K and Rohozi\'{n}ski S~G 1976 {\em Nucl. Phys.
  A\/} {\bf 274} 151

\bibitem{1999LI38}
Libert J, Girod M and Delaroche J~P 1999 {\em Phys. Rev. C\/} {\bf 60} 054301

\bibitem{2010DE02}
Delaroche J~P, Girod M, Libert J, Goutte H, Hilaire S, Peru S, Pillet N and
  Bertsch G~F 2010 {\em Phys. Rev. C\/} {\bf 81} 014303

\bibitem{2004PR03}
Pr\'{o}chniak L and Ring P 2004 {\em Int. J. Mod. Phys. E\/} {\bf 13} 217

\bibitem{2009NI04}
Niksic T, Li Z~P, Vretenar D, Pr\'{o}chniak L, Meng J and Ring P 2009 {\em Phys.
  Rev. C\/} {\bf 79} 034303

\bibitem{x85eis}
Eisenberg J and Greiner W 1985 {\em Nuclear Theory Vol. 1: Nuclear Models\/}
  3rd ed (North Holland Physics Publ.)

\bibitem{x87iac}
Iachello F and Arima A 1975 {\em The Interacting Boson Model\/} (Cambridge
  University Press)

\bibitem{2011BA45}
Baran A, Sheikh J~A, Dobaczewski J, Nazarewicz W and Staszczak A 2011 {\em
  Phys. Rev. C\/} {\bf 84} 054321

\bibitem{1999YU09}
Yuldashbaeva E~K, Libert J, Quentin P and Girod M 1999 {\em Phys. Lett. B\/}
  {\bf 461} 1

\bibitem{xx13BO01}
{Bogner} S, {Bulgac} A, {Carlson} J, {Engel} J, {Fann} G, {Furnstahl} R~J,
  {Gandolfi} S, {Hagen} G, {Horoi} M, {Johnson} C, {Kortelainen} M, {Lusk} E,
  {Maris} P, {Nam} H, {Navratil} P, {Nazarewicz} W, {Ng} E, {Nobre} G~P~A,
  {Ormand} E, {Papenbrock} T, {Pei} J, {Pieper} S~C, {Quaglioni} S, {Roche}
  K~J, {Sarich} J, {Schunck} N, {Sosonkina} M, {Terasaki} J, {Thompson} I,
  {Vary} J~P and {Wild} S~M 2013 {\em Comp. Phys. Comm.\/} {\bf 184} 2235--2250

\bibitem{2003BE12}
Bender M, Heenen P~H and Reinhard P~G 2003 {\em Rev. Mod. Phys.\/} {\bf 75} 121

\bibitem{1995RE03}
Reinhard P~G and Flocard H 1995 {\em Nucl. Phys. A\/} {\bf 584} 467

\bibitem{2000BE32}
Bender M, Rutz K, Reinhard P~G and Maruhn J~A 2000 {\em Eur. Phys. J. A\/} {\bf
  8} 59

\bibitem{2012KO06}
Kortelainen M, McDonnell J, Nazarewicz W, Reinhard P~G, Sarich J, Schunck N,
  Stoitsov M~V and Wild S~M 2012 {\em Phys. Rev. C\/} {\bf 85} 024304

\bibitem{2014KO13}
Kortelainen M, McDonnell J, Nazarewicz W, Olsen E, Reinhard P~G, Sarich J,
  Schunck N, Wild S~M, Davesne D, Erler J and Pastore A 2014 {\em Phys. Rev.
  C\/} {\bf 89} 054314

\bibitem{xx12SC01}
{Schunck} N, {Dobaczewski} J, {McDonnell} J, {Satu{\l}a} W, {Sheikh} J~A,
  {Staszczak} A, {Stoitsov} M and {Toivanen} P 2012 {\em Comp. Phys. Comm.\/}
  {\bf 183} 166--192

\bibitem{2007PR07}
Pr\'{o}chniak L 2007 {\em Int. J. Mod. Phys. E\/} {\bf 16} 352

\bibitem{2010LI09}
Li Z~P, Niksic T, Vretenar D and Meng J 2010 {\em Phys. Rev. C\/} {\bf 81}
  034316

\bibitem{nndc0914}
Data extracted using the NNDC On-Line Data Service from the ENSDF database,
  file revised as of September 2014.

\bibitem{2012WR03}
Wrzosek-Lipska K, Pr\'{o}chniak L, Zieli\'{n}ska M, Srebrny J, Hady\'{n}ska-Klek K,
  Iwanicki J, Kisieli\'{n}ski M, Kowalczyk M, Napiorkowski P~J, Pietak D and
  Czosnyka T 2012 {\em Phys. Rev. C\/} {\bf 86} 064305

\bibitem{1993SR01}
Srebrny J, Czosnyka T, Karczmarczyk W, Napiorkowski P, Droste C, Wollersheim
  H~J, Emling H, Grein H, Kulessa R, Cline D and Fahlander C 1993 {\em Nucl.
  Phys. A\/} {\bf 557} 663c

\bibitem{2009CO24}
Coquard L, Pietralla N, Ahn T, Rainovski G, Bettermann L, Carpenter M~P,
  Janssens R~V~F, Leske J, Lister C~J, Moller O, Rother W, Werner V and Zhu S
  2009 {\em Phys. Rev. C\/} {\bf 80} 061304

\end{thebibliography}
\end{document}